\documentclass[structabstract]{aa}  
\usepackage{graphicx}
\usepackage{natbib}
\usepackage{txfonts}
\begin{document}
   \title{Observations of the recurrent M31 transient XMMU~J004215.8+411924 with Swift, Chandra, HST and Einstein}
   \titlerunning{M31 recurrent transient XMMU J004215.8+411924}

   \author{R. Barnard
          \inst{1}
          \and
          M. Garcia\inst{1}
	  \and
          S. Murray\inst{1}
	  \and
	  N. Nooraee\inst{2}
	  \and
	  W. Pietsch\inst{3}
          }

   \institute{Harvard-Smithsonian Center for Astrophysics, 60 Garden Street, Cambridge MA 02138\\
              \email{rbarnard@head.cfa.harvard.edu}
         \and
             Dublin Institute for Advanced Studies, Dublin,  Republic of Ireland
	 \and
	     Max-Planck-Institut f\"{u}r extraterrestrische Physik, Garching, Germany\\
             }
    \authorrunning{R. Barnard et  al.}
   \date{}

% \abstract{}{}{}{}{} 
% 5 {} token are mandatory
 
  \abstract
   {The transient X-ray source XMMU J004215.8+411924 within M31   was found to be in outburst again in the 2010 May 27 Chandra observation. We present results from our four Chandra and seven Swift observations that covered this outburst.  }
   {X-ray transient behaviour is generally caused by one of two things:  mass accretion from a high mass companion during some restricted phase range in the orbital cycle,   or disc instability in a low mass system. We aim to exploit Einstein, HST, Chandra and Swift observations to determine the nature of XMMU J004215.8+411924.}
   {We model the 2010 May spectrum, and use the results to convert from intensity to counts in the fainter Chandra observations, as well as the Swift observations; these data are used to create a lightcurve. We also estimate the flux in the 1979 January 13 Einstein observation. Additionally, we search for an optical counterpart in HST data.}
   {Our best X-ray positions from the 2006 and 2010 outbursts are 0.3$''$ apart, and 1.6$''$ from the Einstein source; these outbursts are likely to come from the same star system. We see no evidence for an optical counterpart with $m_{\rm B}$ $\la$25.5; this new limit is 3.5 magnitudes fainter than the existing one. Furthermore, we see no V band counterpart with $m_{\rm V}$ $\la$26.  The local absorption is $\sim$7 times higher than the Galactic line-of-sight, and provides $\sim$2 magnitudes of extinction in the V band. Hence $M_{\rm V}$ $\ga -$0.5.   Fits to the X-ray emission spectrum suggest a black hole primary.}
   {We find that XMMU J004215.8+411924 is most likely to be a transient LMXB, rather than a HMXB as originaly proposed. The nature of the primary is unclear, although we argue that a black hole is likely.}
   \maketitle
%
%________________________________________________________________

\section{Introduction}

 The bulge region of M31, the nearest spiral galaxy neighbour, is one of the best laboratories in the Universe for studying X-ray binaries. Accordingly, it has been observed hundreds of times by various X-ray observatories over the past 30 years.  In the last $\sim$10 years alone, it has been observed 120 times by Chandra, 90 times by Swift and 31 times by XMM-Newton. Most of the Chandra and XMM-Newton obbservations were short, monitoring observations looking for transient X-ray sources, while the Swift observations generally followed these transients \citep[see e.g.][]{will05,williams06}. 

XMMU J004215.8+411924 was identified as a new X-ray transient in the 2006, August 9 observation of M31 \citep{haberl06}, with a positional uncertainty of 2$''$. \citet{haberl06} found that the  emission spectrum was well described by a power law with photon index 1.57, suffering absorption equivalent to 4.2$\times10^{21}$ H atom cm$^{-2}$; the unabsorbed 0.5--10 keV luminosity was  9.1$\times10^{37}$ erg s$^{-1}$, assuming a distance of 780 kpc.   The follow-up Swift observation made on 2006 September 1 revealed a UV counterpart within the X-ray error circle, leading \citet{haberl06} to identify XMMU J004215.8+411924 as a high mass X-ray binary (HMXB). A further Swift observation on 2006 September 11 yielded only three photons from XMMU J004215.8+411924 \citep{pietsch06}; this corresponded to a 0.5--10 keV luminosity of $<5\times10^{36}$ erg s$^{-1}$ when assuming the above  emission model.

\citet{galache06} subsequently reported a Chandra detection of XMMU J004215.8+411924 in a July 31  observation, finding the 0.9--6 keV spectrum to be well modeled by a power law model with photon index $\sim$1.8, with absorption equivalent to 4.4$\times$10$^{21}$ H atom cm$^{-2}$. They found the 0.5--10 keV luminosity to be 1.1$\times$10$^{38}$ erg s$^{-1}$.

\citet{voss08} examined the Chandra, Swift and XMM-Newton observations of three transients in M31, including XMMU J004215.8+411924.  The June 2 XMM-Newton observation  made no firm detection of J004215.8+411924;  hence \citet{voss08} constrainted the outburst duration to 40--79 days. They refined the source position by registering the Chandra data with the 2MASS catalogue of \citet{skrutskie2006}. They obtained RA(J2000) = 00:42:16.1, Dec(J2000) = +41:19:26.7, with a 1$\sigma$ uncertainty of 0.5$''$. This new position was $\sim 4''$ from the counterpart identified by \citet{haberl06}, allowing \citet{voss08} to reject this association. They searched for a counterpart in the local group galaxy survey (LGGS) images provided by \citet{massey06}, and found nothing with V $\la$ 22; using a distance modulus of 24.46 and 0.4 mag of extinction, they could not rule out a Be companion star.

In \citet{nooraee10} we reported a new outburst in the 2010 May 27 Chandra observation  within 0.5$''$ of the position of XMMU J004215.8+411924, and identified it as a recurring transient. In this paper we present detailed analysis of the seven Swift and four Chandra observations made between 2010 March 5 and 2010 July 20.  We also search for a counterpart in the 2006 August 27 HST observation, and present evidence for a previous outburst  in 1979 detected with Einstein.

We discuss the observations and data analysis in Section 2. We then present our results in Section 3 and  discuss in Section 4 whether XMMU J004215.8+411924 is a  HMXB displaying orbital variability,  or a transient low mass X-ray binary (LMXB) with unstable disc accretion.

%__________________________________________________________________

%                                   Observations and data analysis
%------------------------------------------------------------------
\section{Observations and data analysis}
\label{obs}

A journal of observations is provided in Table~\ref{journ}.

%_____________________________________________________________
%                                             Simple A&A Table
%_____________________________________________________________
%
\begin{table}
\caption{Journal of observations. For each observation we give the instrument,  observation number, date, exposure time and number of net source photons}             % title of Table
\label{journ}      % is used to refer this table in the text
                         % used for centering table
\renewcommand{\tabcolsep}{3pt}
\begin{tabular}{c c c c c}        % centered columns (4 columns)
\hline\hline                 % inserts double horizontal lines
Instrument & Obs & Date & Exposure & Net counts \\    % table heading 
\hline                        % inserts single horizontal line
   Einstein HRI & 579 & 1979-01-13 & 29 ks & 29 \\   
   Chandra ACIS-I & 7139 & 2006-07-31 & 4 ks      & 433 \\
   Chandra ACIS-I & 11279 & 2010-03-05 & 4 ks  & 5 \\
   Chandra ACIS-I & 11838 & 2010-05-27 & 4 ks & 279\\ 
   Swift XRT& 0031255012 & 2010-06-06 &4 ks  & 44\\
   Swift XRT& 0031255013 & 2010-06-09 &4 ks & 37\\
   Swift XRT& 0031255014 & 2010-06-12 &4 ks & 24\\
   Swift XRT& 0031255015 & 2010-06-15 & 1.7 ks & 9\\
   Swift XRT& 0031255016 & 2010-06-18 & 3 ks  & 6$^{a}$\\
   Chandra ACIS-I& 11839 & 2010-06-23 & 4 ks & 27\\ 
   Swift XRT& 0031255018 & 2010-06-24 &4 ks  &1$^{a}$\\
   Chandra ACIS-I& 11840 & 2010-07-20 & 4 ks & $\sim$0$^{a}$\\
\hline                                   %inserts single line
\end{tabular}
\\$^{a}$ Not a secure detection. Derived luminosities are 3$\sigma$ upper limits.
\end{table}

\subsection{Analysis of Chandra data}

The 2006 July 31 Chandra observation (7139) of   XMMU J004215.8+41192 has already been discussed by \citet{voss08}. However, we registered the image to the LGGS B band image of M31 Field 6, which has positional uncertainties of 0.25$''$, using globular clusters (GCs)  in the Revised Bologna Catalogue V4 \citep{galleti04, galleti06, galleti07, galleti09} that were X-ray bright. To do this, we used the {\sc iraf} tool {\sc imcentroid} to determine the position of each cluster in the Chandra and LGGS observations in image coordinates, then checked the positional uncertainties of each GC.  We then used {\sc xy2sky} to get the sky coordinates for each GC in the X-ray and LGGS images.  This allowed us to map the LGGS coordinates to the X-ray positions with {\sc ccmap}.  The position of XMMU J004215.8+41192 was determined in the corrected image using {\sc imcentroid}, and the positional uncertainties were combined with the 0.25$''$ of the LGGS coordinate system.

For the 2010 May 27 Chandra observation (11838), we located XMMU J004215.8+41192 as above. In addition, we extracted source and background spectra, with related products, using {\sc ciao} ver 4.2. These spectra were analysed with {\sc xspec ver 12.6}.  XMMU J004215.8+41192 was too faint for spectral modelling in the 2010 March 5 , 2010 June 24 and 2010, July 20 observations; instead, we derived a conversion from intensity to flux for each observation  using the best fit emission model from Obs 11838, as described below.

\subsection{Analysis of Swift data}

We were awarded seven observations of M31 with Swift during 2010,  June at three day intervals to monitor the transient CXOM31 J004253.1+4122 \citep{henze09}; results from that transient will be presented by  Nooraee et al. (in prep).   This work uses data from the X-ray Telescope (XRT), taken in photon counting (pc) mode.  Unfortunately, the 2010 June 21 observation had an  exposure time of just 35s in pc mode, so this observation was ignored.   XMMU J004215.8+411924 was at a  large off-axis angle that varied between observations ($\sim$7--10$'$).

The Swift data were analysed with {\sc xselect} version 2.4a. We were unable to use the recommended source extraction radius of 47$''$ (containing $\sim$ 90\% of the source photons) due to crowding. Instead we used a circular source extraction region with radius 20$''$, and a concentric,  annular background region with inner radius 20$''$ and outer radius 28.28$''$.  Lightcurve analysis was not productive, since the Swift data are obtained in chunks of a few hundred seconds, and very few photons were collected each time.

Since there were too few photons for spectral analysis in each Swift observation, we converted from background-subtracted intensity to 0.5--10 keV flux by calculating the flux equivalent to 1 count s$^{-1}$ assuming the best fit model derived from our analysis of the  Obs 11838 Chandra data. To do this we obtained source and background spectra for each observation, and created an ancillary response file using the tool {\sc xrtmkarf}; the appropriate canned response matrix was used. For each observation we loaded these files into {\sc xspec} version 12.6 and normalised the best fit emission model to give 1 count s$^{-1}$.  This enabled us to obtain the unabsorbed 0.5-10 keV flux equivalent to 1 count s$^{-1}$, which we refer to as the conversion factor.

\subsection{Analysis of HST data}
We triggered two HST observations after the 2006 outburst, designed to observed the counterpart in the on phase  phase and off phase. The first observation was made with the ACIS/WFC on 2006 August 27 using the F435W filter. Unfortunately, the ACS was not operational during our second observation in 2007 July, and the observation was made with WFPC/FIX-1 using the F439W filter. However, the position of XMMU J004215.8+411924 was serendipitously covered by an ACS/WFC1 observation in 2004, October using the F555W filter.

We obtained the drizzled images of these observations from the Hubble legacy archive, and registered them with the  B image of M31 Field 6 in the LGGS; we used  stars that were visible in both images, but not so bright that the centroid determination had large uncertainties.

%                                 RESULTS
%------------------------------------------------------------------------------
\section{Results}
\label{results}
\subsection{Fitting the 2010 May Chandra spectrum}

The 2010 May observation of  XMMU J004215.8+411924 yielded 279 net source counts. We grouped the source spectrum to get a minimum of 15 counts per bin. The best fit power law emission model had a photon index of 1.8$\pm$0.5, with absorption equivalent to 6$\pm 3\times10^{21}$ H atom cm$^{-2}$; $\chi^2$/dof =7/15. These parameters are entirely consistent with those found by \citet{voss08}.

\begin{figure}
   \resizebox{\hsize}{!}{\includegraphics[scale=0.3,angle=270]{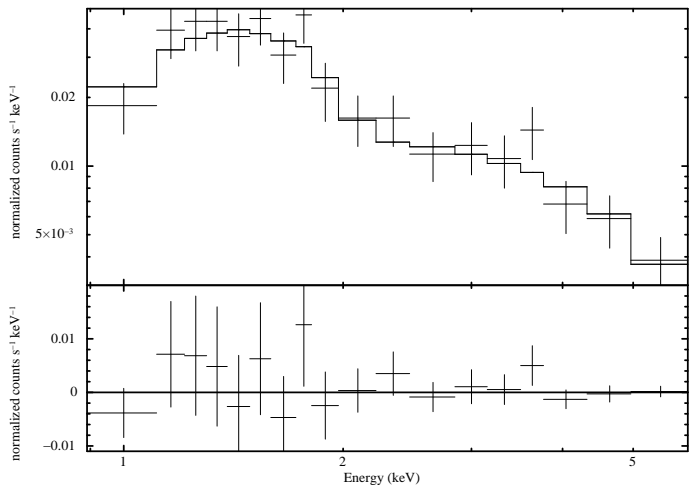}}
      \caption{ Best fit power law model to the Obs 11838 Chandra spectrum, with residuals; $\chi^2$/dof = 7/16. $N_{\rm H}$ was fixed to 4.2$\times 10^{21}$ atom cm$^{-2}$.  }  \label{specfig}
   \end{figure}

The XMM-Newton observation analysed by \citet{voss08} yielded the tightest constraints on the absorption (4.2$\pm0.5\times 10^{21}$ H atom cm$^{-2}$); we therefore fixed the absorption to 4.2$\times10^{21}$ H atom cm$^{-2}$ for our Chandra spectrum. This gave a best fit photon index of 1.6$\pm$0.2, and $\chi^2$/dof =8/16. This yielded an unabsorbed 0.5-10 keV luminosity of 1.2$\pm0.3\times 10^{38}$ erg s$^{-1}$.  This fit is presented in Fig.~\ref{specfig}.

%                                                One column figure
%----------------------------------------------------------specfig
%
%______________________________________________________________

 We also fitted the spectrum with a disk blackbody model, since black holes in outburst are often in the high soft state, with thermal emission spectra \citep[see e.g.][ and references within]{mr06}. The best fit spectrum had an inner disc temperature of 1.4$\pm$0.3 keV; $N_{\rm H}$ was fixed to 4.2$\times 10^{21}$ atom cm$^{-2}$ as before; $\chi^2$/dof = 10/16. This temperature is somewhat high for a high state black hole binary \citep{mr06}, and the fit is rather worse than the power law fit. Hence we prefer the power law fit, but cannot rule out a thermal fit.

A neutron star accreting at $\sim 10^{38}$ erg s$^{-1}$ has a two component emission spectrum, often characterised as a blackbody and a power law, with the blackbody contributing $\sim$10-50\% of the luminosity \citep[see e.g][]{cbc95, church02}. Hence, we fitted the spectrum with a blackbody + power law model. The best fit gave kT $\sim$ 1 keV, and a power law slope 1.6$\pm$ 0.6; $\chi^2$/dof =7/14; however, the blackbody component was not well constrained, and only contributed $\sim$5\% of the flux, significantly less than expected for a neutron star system. We infer from the lack of a strong thermal component that the primary is more likely to be a black hole than a neutron star; however, we cannot rule out a neutron star primary. 

\subsection{The X-ray lightcurve}

We obtained the conversions from intensity to 0.5--10 keV flux for each Swift observation and the 2010 June Chandra observation, assuming a power law with photon index 1.6, with $N_{\rm H}$ = 4.2$\times 10^{21}$  atom cm$^{-2}$. The conversion factor varied by up to 25\% between observations, due to changes in off-axis angle.

\begin{figure}
   \resizebox{\hsize}{!}{\includegraphics[scale=0.3,angle=0]{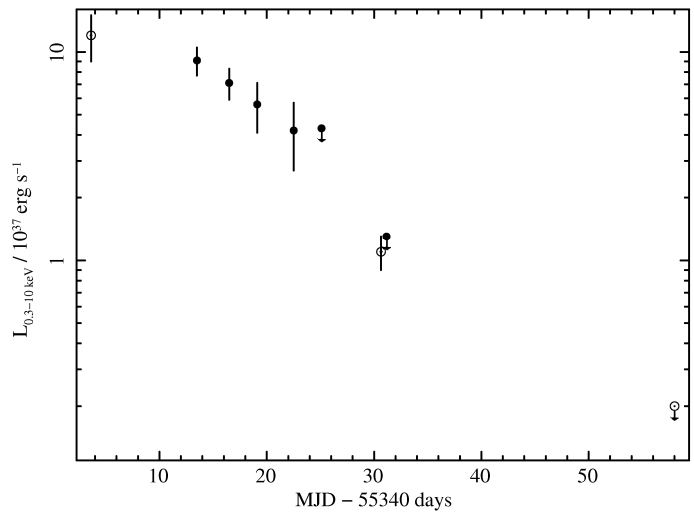}}
      \caption{0.3-10 keV lightcurve of XMMU J004215.8+411924 from Swift (filled circles) and Chandra (hollow circles). Arrows represent 3$\sigma$ upper limits. The y axis is log scaled for clarity.  }  \label{lc_fig}
   \end{figure}

We present the lightcurve of XMMU J004215.8+411924 covering the outburst from 2010 May 27  to 2010 July 20 in Fig.~\ref{lc_fig}. Each point is plotted at the midpoint of the  observation;  the points from the June 18 and June 24 Swift observations, and the July 20 Chandra observation, are 3$\sigma$ upper limits.  The 2010 July Chandra observation yielded a 3$\sigma$ upper limit of 2$\times 10^{36}$ erg s$^{-1}$, suggesting that  XMMU J004215.8+411924 may have  returned to its pre-outburst level. 

 The deepest observation  made of XMMU J004215.8+411924 is Chandra Obs 1575, with a $\sim$40 ks duration. The 3$\sigma$ upper luminosity limit is 6$\times 10^{35}$ erg s$^{-1}$. Hence, the  variation in luminosity is a factor $\ga$200 between the outburst peak and quiescence. 

 XMMU J004215.8+411924 was observed by Chandra three times in 2010 prior to the detection of the outburst, in January, February and March; no observation was made in April, as M31 was behind the sun. We found no strong detections in any of these observations; the 3$\sigma$ upper limit for the March observation was $\sim$2$\times 10^{36}$ erg s$^{-1}$. Hence the 2010 outburst lasted at least 30 days, but not more than 140 days. The outburst appears to be of similar duration to the one in 2006.

 We have found no evidence of additional outbursts in the 120 Chandra observations and 90 Swift observations of XMMU J004215.8+411924. 

%                                                One column figure
%-----------------------------------------------------------rtslc
%
%______________________________________________________________

\subsection{The X-ray position of XMMU J004215.8+411924}

For the 2006 July Chandra  observation,  we chose 12 bright X-ray sources associated with GCs for the registration. Five GCs had unacceptably large uncertainties  in their X-ray positions ($>$0.2$''$), and these were removed from the registration process. Additionally, one of the GCs showed an unusually large discrepency between the X-ray and optical positions, so it was discarded and a new solution was found; the final r.m.s. offset between Chandra and LGGS was 0.07$''$ in RA and 0.12$''$ in Dec. Our best registration solution yielded RA = 00:42:16.063 Dec = +41:19:26.73, with 0.17$''$ uncertainty in RA and 0.2$''$ uncertainty in Dec.  Combining these with the position uncertainties of the LGGS images gives an error circle with 0.3$''$ radius.

Our final registration of the 2010 May Chandra observation utilised five GCs; the final r.m.s. offset between Chandra and LGGS was 0.19$''$ in RA and 0.16$''$ in Dec. Our best location in this observation was RA = 00:42:16.037, Dec = +41:19:26.63, with 0.17$''$ uncertainty in RA and 0.28$''$ uncertainty in Dec. Combining these with the LGGS uncertainties results in a 0.3$''\times0.4''$ ellipse.
The best fit positions in the two observations are offset by 0.3$''$. Therefore it is most likely that both X-ray outbursts come from the same source.

\subsection{The Einstein outburst}

We note that our 2010 position is 1.6$''$ from an X-ray source detected in the Einstein HRI $\sim$30 years previously  \citep[source 17 in][]{crampton84}; the uncertainty in Einstein position is $\sim$10$''$, and includes no known X-ray sources other than XMMU J004215.8+41192 This source was not detected by \citet{tf91}, who used a source detection radius of 6.7$''$ for the Einstein HRC; however, XMMU J004215.8+41192 was observed at a high off-axis angle, and the photons were spread over a wider area than the detection cell. 

Extracting a circle with 20$''$ radius around the position of XMMU J004215.8+41192 yields 76 counts, while a nearby source free region of the same size  yields 47 counts. The net exposure time for the observation is 28564 s. We calculated a conversion from 0.2--4.0 keV intensity in the HRC to 0.5--10keV unabsorbed flux using {\sc WebPIMMS}: 1 count s$^{-1}$ = 7.1$\times 10^{-10}$ erg cm$^{-2}$ s$^{-1}$. Hence we estimate the 0.5--10 keV luminosity of   XMMU J004215.8+41192 to have been $\sim 5 \times 10^{37}$ erg s$^{-1}$ in the 1979 January 13 observation. 

%                                                One column figure
%-----------------------------------------------------Einsteinim
%   \begin{figure}
%   \includegraphics[scale=0.35,angle=0]{rts_e.eps}
%      \caption{0.5-10 keV lightcurve of XMMU J004215.8+411924 from Swift (circles) and Chandra (stars).  }  \label{ein_fig}
%   \end{figure}
%
%______________________________________________________________

\subsection{The search for counterparts}

The 2006, August HST observation was registered to the LGGS using seven stars, with a r.m.s  offset of 0.02$''$ in both RA and Dec. The 2005, May HST observations was registered with 5 stars, resulting in r.m.s. offsets of 0.019$''$ and 0.05$''$ in RA and Dec respectively. 

\begin{figure}
   \includegraphics[scale=0.22,angle=0]{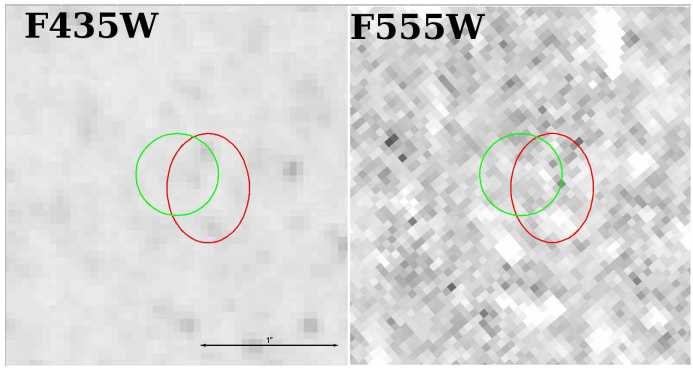}
      \caption{Details of HST images of the environs of XMMU J004215.8+411924 from 2006, August, i.e. during outburst (left) and 2004, October, in quiescence (right). Uncertainties in the X-ray position are represented  by a circle for the 2006 Chandra observation and by an ellipse for the 2010 May Chandra observation.}  \label{hst_fig}
   \end{figure}

In Fig~\ref{hst_fig} we present details of the HST images from 2006, August (left) and 2005, May (right). These images are superposed with a circle that represents the X-ray position from the 2006, July Chandra observation, and an ellipse to represent the 2010, May position.   We find no compelling evidence for a counterpart during outburst down to $m_{\rm B}$ $\sim$25.5, or $m_{\rm V}$ $\sim$26, let alone evidence for variability that would confirm the association. 

 At a distance of 780 kpc, M31 has a distance modulus of 24.45 magnitudes. The observed column density is significantly higher than the Galactic line-of-sight absorption (6$\times10^{20}$ H atom cm$^{-2}$); using the empirical relationship obtained by \citet{predehl95}, we can expect $\sim$2 magnitudes of V band  extinction, and $\sim$3 magnitudes in B band. Hence we place upper limits on the counterpart of $M_{\rm B}$ $\ga$ $-$2, and  $M_{\rm V}$ $\ga$ $-$0.5.

%                                                One column figure
%-----------------------------------------------------------hst_fig
%
%______________________________________________________________

%                                                         DISCUSSION
%-------------------------------------------------------------------

\section{Discussion}
\label{discuss}

Two mechanisms could be responsible for the huge variation in mass accretion that resulted in the observed outburst \citep[see e.g.][for a discussion of M31 transients]{williams06}. The system could be a HMXB with a long, eccentric orbital period where accretion is intensified near periastron; this scenario was initially favoured by \citet{haberl06} after mistakenly identifying a counterpart in the Swift UVOT image. Alternatively, the system could be a low mass X-ray binary (LMXB) with an unstable accretion disc that oscillates between a cold  state (quiescence) and a hot, ionised state (outburst), see e.g. \citet[ and references within]{dubus01}. Here we consider the observational constraints on both scenarios.

\subsection{Constraints on a HMXB  system}

The known counterparts of HMXBs in the SMC have apparent $V$ magnitudes  in the range 13 $\la$ $m_{\rm V}$ $\la$ 18, and $B-V$ in the range $-$0.32 $\le$ $B-V$ $\le$ 0.06 \citep[see e.g.][]{coe05,antoniou09}. For a distance of $\sim$60 kpc, this equates to $-$6 $\la$ $M_{\rm V}$  $\la$ $-$1, all brighter than our threshold of $M_{\rm V}$ $\ga$-0.5. It is therefore unlikely that a Be star is hidden by the local absorption.

 Furthermore, variations in accretion rate on the orbital cycle are  of course periodic, and we see no evidence for other outbursts in our $\sim$120 other Chandra monitoring observations. Since the outbursts lasted at least  $\sim$30 days, and the frequency of our monitoring is once per $\sim$30 days, we would expect coverage of other outbursts.

If the outbursts in 1979, 2006 and 2010 were due to periodic accretion near perihelion, then a whole number of orbital cycles every  $\sim$1430 days would be required. Hence,   XMMU J004215.8+41192 would have had an unabsorbed 0.5--10 keV luminosity $\sim$10$^{38}$ erg s$^{-1}$ during the 2006, July 2 XMM-Newton observation. However, \citet{voss08} did not detect XMMU J004215.8+4119 during that observation. 
We therefore conclude that  XMMU J004215.8+411924 is likely to be a LMXB.

\subsection{Constraints on a LMXB  system}

The optical emission of X-ray bright LMXBs is dominated by the accretion disc; since larger discs exist in systems with longer orbital periods, \citet{vanparadijs94} derived an empirical relation between the optical luminosity  ($L_{V}$), the X-ray luminosity ($L_{V}$) and the disc radius ($R$) : $L_{\rm V}$  $\propto$ $L_{\rm X}^{1/2} R$. The range in absolute  magnitudes of the sample was $-$5 $\la$ $M_{\rm V}$ $\la$ 5.

Including the relation between R and orbital period, and defining $\Sigma$ as $\left( L_{\rm X}/ L_{Edd}\right)^{1/2} \left(P/1{\rm hr}\right)^{2/3}$, they further found that the relation
\begin{equation}
M_{\rm V} = 1.57\pm0.24 -2.27\left(\pm0.32\right)\log \Sigma
\end{equation}
produced good results over three orders of magnitude in $\Sigma$.  This relation allows us to estimate the range in orbital period for XMMU J004215.8+411924. Due to the high local absorption, the upper limit to the orbital period is not particularly constraining: $\la$40 hr for a neutron star and $\la$ 130 hr for a black hole,  longer than typical LMXB periods.

 The X-ray spectrum from Observation 11838 is best fitted by  a power law fit with spectral index $\sim$1.6; this is seen in neutron star and black hole binaries at low accretion rates \citep{vdk94}. However, \citet{glad07} showed that neutron star LMXBs don't exhibit this behaviour at 0.01--1000 keV luminosities $\ga$10\% of the Eddington limit.  Neutron star LMXBs at higher luminosities have two component emission spectra, with a thermal component contributing $\sim$10--50\% of the flux \citep{church02}. Since the 0.5--10 keV luminosity of  XMMU J004215.8+411924 is $\sim$70\% of the Eddington limit for a 1.4 M$_\odot$ neutron star, the primary is likely to be a black hole \citep[][]{barnard08}; this conclusion is supported by the lack of a strong thermal component in the two component model. However, the quality of the data prevents us from excluding a neutron star primary. 

\begin{acknowledgements}
 We thank the anonymous referee for their thoughtful comments.  We thank the Swift team for carrying out the Target of Opportunity observations. RB is funded by Chandra grant GO9-0100X and HST grant GO-11013. MRG is partially supported by  NASA grant NAS-03060.  This work has made use of the data obtained
through the High Energy Astrophysics Science Archive Research Center
Online Service, provided by the NASA/Goddard Space Flight Center, and the Hubble Legacy Archive.    
\end{acknowledgements}

\bibliographystyle{aa}
\bibliography{mnrasm31}

\end{document}